\documentclass[preprint,12pt]{elsarticle}
\usepackage{graphicx} % Required for inserting images
\RequirePackage{fix-cm}
\usepackage{graphicx}
\usepackage{braket}
\usepackage{qcircuit}
\usepackage{csquotes}
\usepackage{float}
\usepackage{ragged2e}
\usepackage{multirow}
\usepackage{array}
\usepackage{dcolumn}
\usepackage{hhline}
\usepackage{amssymb}
\usepackage{amsmath}
\usepackage{sidecap,tikz, multirow,subcaption}

\usepackage{xcolor}
\journal{Annals of Physics}

\usepackage{longtable}
\usepackage{amsfonts}

\expandafter\let\csname equation*\endcsname\relax
\expandafter\let\csname endequation*\endcsname\relax

\usepackage[breaklinks=true,colorlinks=true,linkcolor=blue,urlcolor=blue,citecolor=blue]{hyperref}

\begin{document}

\begin{frontmatter}

\title{A maximum concurrence criterion to investigate absolutely maximally entangled states}

\author[inst1]{Subhasish Bag}

\affiliation[inst1]{organization={ Department of Physics},%Department and Organization
            addressline={}, 
            city={Indian Institute of Technology Delhi, Hauz Khas},
            postcode={110016}, 
            state={ New Delhi},
            country={India}}

\author[inst3,inst2]{Ramita Sarkar \corref{cor1}}
\author[inst4,inst2]{Prasanta K. Panigrahi}

\affiliation[inst3]{organization={ Institute of Physics Bhubaneswar},%Department and Organization
             addressline={}, 
            city={Bhubaneswar},
            postcode={751005}, 
            state={Odisha},
            country={India}}
\affiliation[inst4]{organization={ Center for Quantum Science and Technology, Siksha ’O’ Anusandhan University},%Department and Organization
             addressline={}, 
            city={Bhubaneswar},
            postcode={751030}, 
            state={Odisha},
            country={India}}            
\affiliation[inst2]{organization={ Department of Physical Sciences},%Department and Organization
addressline={Indian Institute of Science Education and Research Kolkata}, 
city={Mohanpur},
 postcode={741246}, 
state={West Bengal},
country={India}}
\cortext[cor1]{Corresponding author}
\begin{abstract}
We propose a straightforward method to determine the maximal entanglement of pure states using the criterion of maximal I-concurrence, a measure of entanglement. The square of concurrence for a bipartition $X|X^\prime$ of a pure state is defined as $E^2_{X| X ^\prime}=2[1-tr({\rho_X}^2)]$. From this, we can infer that the concurrence  $E_{X| X ^\prime}$ reaches its maximum when $tr({\rho_X}^2)$ is minimized. Using this approach, we identify numerous Absolutely Maximally Entangled (AME) pure states that exhibit maximal entanglement across all possible bipartitions. Conditions are derived for pure states to achieve maximal mixedness in all bipartitions, revealing that any pure state with an odd number of subsystem coefficients does not meet the AME criterion.  Furthermore, we obtain equal maximal multipartite entangled pure states across all bipartitions using our maximal concurrence criterion.  
\end{abstract}

\begin{highlights}
\item  Concurrence
\item  Entropy
\item  Maximally Entangled States
\item  Absolutely Maximally Entangled (AME) States
\item  Quantum Graph and Hypergraph States

\end{highlights}


\end{frontmatter}

\section{Introduction}
{Quantum entanglement is the foundational concept for various quantum information and communication applications well understood for bipartition systems. Recently, multipartite entanglement is being explored\cite{walter2013entanglement,szalay2015multipartite}, extending our understanding beyond simple bipartite entanglement. Multiparticle entanglement is essential for comprehending potential quantum benefits in information processing\cite{PhysRevA.72.062108} or meteorology\cite{PhysRevLett.126.080502}, error correction\cite{campbell2024series}. When investigating where the multipartite quantum states exhibit the highest entanglement, researchers have harnessed the well-understood domain of bipartite entanglement to scrutinize multipartite states. The term ``maximal multipartite entangled states"\cite{1s} is synonymous with the utmost degree of entanglement for each possible bipartition of parties. An n party quantum state having reductions to $\lfloor n/2 \rfloor$ parties that are maximally mixed \cite{2s,3s}, then these states are known as Absolutely Maximally Entangled (AME) states \cite{4s,5s,6s,1s,8s,9s,10s,11s,12s,13s,14s,15s,16s,17s,18s,19s}. These AME states contribute to the developing field of quantum technology in quantum secret sharing\cite{18s,11s}, quantum teleportation\cite{11s}, quantum error-correcting codes\cite{4s}, and various other applications, contributing to the evolving landscape of quantum technology.}

{Numerous efforts have been dedicated to uncovering Absolutely Maximally Entangled (AME) states, particularly for qubit cases. The quest for AME(n, d) states, where n denotes the number of parties and d represents the dimension\cite{18s}, has been extensively explored and fruitfully realized for $d$ = 2, with various values of n, such as 2, 3, 5, 6, and beyond. These states are often characterized as stabilizer states \cite{4s,5s,6s,1s,8s,9s,10s,11s,12s,13s,14s,15s,16s,17s,18s,19s}; their existence and properties have been established in the literature. However, it is worth noting that AME states are not universal regarding existence with qubit constituents, as exemplified by the absence of four and seven qubit AME states, a conclusion drawn from rigorous investigations\cite{3s,22s}. Nevertheless, researchers have identified specific instances of AME states that have been reported\cite{32s}, opening new avenues in the quest for AME states in higher dimensions. To identify AME states, a range of analytical and numerical approaches have been explored, including techniques based on graph states\cite{19s,21s}, classical maximum distance separable codes\cite{18s,23s}, combinatorial designs\cite{14s,25s}, and more. It's interesting to note that AME graph states\cite{19s} are a special kind of AME states that have been extensively researched \cite{3s,4s}. These multipartite, highly entangled pure states are very promising for quantum protocol development. Interestingly, cluster states, graph states, and hypergraph states have become indispensable tools in this regard. Graph states, characterized by high entanglement, can be mathematically associated with graphs, where vertices and edges correspond to quantum spin systems and Ising interactions, respectively\cite{hein2004multiparty}. These states are versatile and can be mapped onto various regular 2D lattices\cite{briegel2009measurement}. The criterion for the class of 2-colorable graphs is precisely studied, which is particularly useful in the context of entanglement purification \cite{hein2004multiparty}. Hypergraph states\cite{sarkar2021geometry}, an extension of graph states with more than two connected vertices by a single edge\cite{rossi2013quantum}, offer extensive utility in multiple domains\cite{dutta2019permutation}. These states are not only essential in quantum error correction\cite{schlingemann2001quantum,wagner2018analysis} and quantum blockchain\cite{banerjee2020quantum} but have also found applications in neural networks\cite{yang2021representations,sarkar2021phase}. To measure the entanglement of these states, concurrence \cite{wootters1998entanglement} is one of the approaches that vanishes for a separable state \cite{sarkar2021geometry}. Concurrence is first coined by Hill and Wooters, which has been initially depicted for two-qubit states \cite{PhysRevLett.78.5022}, that it has been generalized for higher qubits \cite{bhaskara2017generalized} }. Recently, concurrence has been used as a measure for continuous variable mixed state \cite{swain2022generalized}. Here, we have constructed the maximum concurrence criterion to find the AME states based on the concurrence measure\cite{bhaskara2017generalized}, that is, a geometrical study based on wedge product. Using this maximum concurrence concept, we have shown the existence of a large number of pure AME states for some qubits.

The paper is organized as follows: in Sec-\ref{sec1}, we present the basics of Concurrence and AME states. In Sec-\ref{sec2}, the maximum concurrence criterion for three, four and five qubits have been explored. The large number of AME states for these specific qubits is examined in Sec-\ref{sec3}. In this section, we have shown different classes of states satisfying absolutely maximally entangled criteria. One particular entangled class, known as Equal Maximally Entangled (EME) states, has been examined in Sec-\ref{sec4}. Finally, in Sec-\ref{sec5}, we summarize our main results and discuss the future scope of this work.

\section{Basics of Concurrence and AME states} \label{sec1}

\subsection{Concurrence:}
 Concurrence is an important measure of entanglement \cite{wootters1998entanglement}, strictly positive for
	entangled states and vanishing for all separable states. This measure quantifies the entire entanglement, providing necessary and sufficient conditions for separability.
	The square of concurrence for a bipartition $X|X^\prime$ is defined as \cite{sarkar2021geometry,dutta2019permutation},
	\begin{equation}\label{1}
		E^2_{X| X ^\prime}=2[1-tr({\rho_X}^2)]
	\end{equation}
	where $\rho_X$ is the reduced density matrix of the sub-system $X$.\\

\subsection{AME States:}
The $n$-qubit $d$-dimensional AME$(n,d)$\cite{4s,5s,6s,1s,8s,9s,10s,11s,12s,13s,14s,15s,16s,17s,18s,19s} states have the von Neumann entropy corresponding to the density matrix $\rho_A$ as
\begin{equation}\label{eq:2}
    S(\rho_A)=Tr[\rho_A{log}_d\rho_A]
\end{equation} being equal to $\lfloor n/d \rfloor$ for the maximally mixed criterion.
Here $\lfloor . \rfloor$ is the floor function. The AME state is also a $k$-uniform state. A $k$-uniform state of an $N$-party system has a maximally mixed state for the $k$ subsystem by tracing out the arbitrary $N-k$ subsystems. Here $k$, the integer can not be greater than $N/2$, and $k=\lfloor n/2 \rfloor$ represents an AME state \cite{11s,14s,19s}. If $k$ is greater than 1, a $k$-uniform state is also a $k'$ uniform state for any $0<k'<k$.

\subsection{Connection between Concurrence and AME States:}
To define a AME state, the von Neumann entropy given by the equation (\ref{eq:2}) provides a limiting condition. Concurrence and linear entropy are known to be related\cite{raggio1995properties,simon2020entropy,mishra2024geometric}. 

Here, the linear entropy(Quantum Tsallis 2-entropy) for a bipartition $X|X^\prime$ has been expressed as\cite{raggio1995properties,simon2020entropy} 
\begin{equation}
    S_L(\rho_X)=\frac{1}{2}E^2_{X| X ^\prime}=[1-tr({\rho_X}^2)]
\end{equation}
The highest concurrence also shows the maximum mixedness for quantum states along the bipartition. We will see later how the highest concurrence is related to the AME states (related to von Neumann entropy) in the section (\ref{sec3}).

\section{Maximum Entangled Criterion} \label{sec2}
In this section, we will explore the conditions of maximally entangled states. From the expression (\ref{1}), one observes that $tr{(\rho_{X})^2}$ will be the minimum for the maximally entangled state. Hence, it is evident that the maximum value for any bipartition will be less than $\sqrt{2}$ as $tr(\rho_X)^2\ne{0}$. The consequence of this is explained below for three, four and five qubit states.

\subsection{Three qubit Maximally Entangled States:}
Suppose we have a three-qubit pure state labeled by $A$, $B$, and $C$ and represented as
\begin{equation}\label{3}
    \ket{\psi}=[a\ket{000}+b\ket{001}+c\ket{010}+d\ket{011}+e\ket{100}+f\ket{101}+g\ket{110}+h\ket{111}]
\end{equation}
In addition, we define a term called `complementary coefficient' in terms of the complementary bit representation of the subsystems. According to that, the coefficients $a$ and $h$ are complementary coefficients as they are the coefficients of $\ket{000}$ and $\ket{111}$, respectively. Similarly, the complementary coefficients of $b,c$, and $d$ are $g,f,$ and $e$, respectively. The bipartite cases of this state (\ref{3}) are  $A|BC$, $B|AC$, $C|AB$ and the total entanglement is given as
\begin{equation}
    E=E_A+E_B+E_C
\end{equation}
where $E$ is the total concurrence and $E_A$, $E_B$, and $E_C$ are the concurrences of the bipartitions corresponding to the substitutions $A$, $B$, and $C$ respectively. \\
Considering the pure state (\ref{3}), we derive the $Tr(\rho_X)^2$ (where $x$=$A$, $B$, $C$; and `-' is used to represent the complex conjugate of the coefficients) as
\begin{equation}\label{4}
\begin{aligned}
    & Tr(\rho_A)^2\\
    &   =(|a|^2+|b|^2+|c|^2+|d|^2)^2+2(a\Bar{e}+b\Bar{f}+c\Bar{g}+d\Bar{h})(e\Bar{a}+f\Bar{b}+g\Bar{c}+h\Bar{d})\\
    & +(|e|^2+|f|^2+|g|^2+|h|^2)^2 \\
                  & =(|a|^2+|b|^2+|c|^2+|d|^2)^2+2|a\Bar{e}+b\Bar{f}+c\Bar{g}+d\Bar{h}|^2\\
                  &+(|e|^2+|f|^2+|g|^2+|h|^2)^2
\end{aligned}
\end{equation}
\begin{equation}\label{5}
    \begin{aligned}
        & Tr(\rho_B)^2\\
        &  = (|a|^2+|b|^2+|e|^2+|f|^2)^2+2(a\Bar{c}+b\Bar{d}+e\Bar{g}+f\Bar{h})(c\Bar{a}+d\Bar{b}+g\Bar{e}+h\Bar{f})\\
        & +(|c|^2+|d|^2+|g|^2+|h|^2)^2\\
                      & =(|a|^2+|b|^2+|e|^2+|f|^2)^2+2|a\Bar{c}+b\Bar{d}+e\Bar{g}+f\Bar{h}|^2\\
                      & +(|c|^2+|d|^2+|g|^2+|h|^2)^2
    \end{aligned}
\end{equation}
\begin{equation}\label{6}
    \begin{aligned}
        &Tr(\rho_C)^2\\
        & =(|a|^2+|c|^2+|e|^2+|g|^2)^2+2(a\Bar{b}+c\Bar{d}+e\Bar{f}+g\Bar{h})(b\Bar{a}+d\Bar{c}+f\Bar{e}+h\Bar{g})\\
        & +(|b|^2+|d|^2+|f|^2+|h|^2)^2\\
                     & = (|a|^2+|c|^2+|e|^2+|g|^2)^2+2|a\Bar{b}+c\Bar{d}+e\Bar{f}+g\Bar{h}|^2\\
                     &+(|b|^2+|d|^2+|f|^2+|h|^2)^2
    \end{aligned}
\end{equation}

Again, if we express a three-qubit pure state as:
\begin{equation}
    \ket{\psi}=\sum_{i,j,k=0}^1a_{ijk}\ket{ijk}
\end{equation}
and considering $a_{ijk}$ as $a_l$, where $l$ varies from 1 to 8; the equations (\ref{4}), (\ref{5}) and (\ref{6}) can be expressed as
\begin{equation}
    Tr(\rho_A)^2=(\sum_{i=1}^{4}|a_i|^2)^2+2\sum_{i=1}^{4}(a_i\Bar{a}_{i+4})\sum_{i=1}^{4}(a_{i+4}\Bar{a}_{i})+(\sum_{i=1}^{4}|a_{i+4}|^2)^2
\end{equation}
\begin{equation}
    \begin{aligned}
   & Tr(\rho_B)^2=\\
   &(\sum_{i=1}^{2}(|a_i|^2+|a_{i+4}|^2))^2 
    +2\sum_{i=1}^{2}(a_i\Bar{a}_{i+2}+a_{i+4}\Bar{a}_{i+6})\sum_{i=1}^{2}(a_{i+2}\Bar{a}_{i}+a_{i+6}\Bar{a}_{i+4})\\
    & +(\sum_{i=1}^{2}(|a_{i+2}|^2+|a_{i+6}|^2))^2
    \end{aligned}
\end{equation}
\begin{equation}
    Tr(\rho_C)^2=(\sum_{i=1}^{4}|a_{2i-1}|^2)^2+2\sum_{i=1}^{4}(a_{2i-1}\Bar{a}_{2i})\sum_{i=1}^{4}(a_{2i}\Bar{a}_{2i-1})+(\sum_{i=1}^{4}|a_{2i}|^2)^2
\end{equation}
We assume the three-qubit GHZ state as a reference state (a maximally entangled state), the corresponding $Tr(\rho_A)^2$, $Tr(\rho_B)^2$, $Tr(\rho_C)^2$ are 1/2 (where $a=h=\frac{1}{\sqrt{2}}$). To get a higher entangled state than the GHZ state, we must have the value of this trace operation less than 1/2. We will explore whether a three-qubit state has higher concurrence than a GHZ state. From the expressions of (\ref{4}, \ref{5}, \ref{6}), it is evident that this will be possible if the combination of the coefficients of a pure state is in such a way that $Tr(\rho_X)^2$ is less than $\frac{1}{2}$. It is already established that one can not achieve the concurrence for a three-qubit pure entangled state more than the GHZ state as this single bipartition is maximally mixed, suggesting that the maximum concurrence for any bipartition (single-cut) is 1.  However, one can still have the same entanglement(concurrence) as the GHZ state. Our task is to find the different cases for a three-qubit pure state having the same entanglement as a three-qubit GHZ state. There are different possible cases discussed in the subsections below.

\subsubsection{States with two real coefficients:}
 We have found that if only the complementary coefficient exists in the equation (\ref{3}), e.g. ($a,h; b,g; c,f; d,e$), whatever the sign, we have a maximally entangled state. Ex- a three-qubit GHZ state having $a=h=+\frac{1}{\sqrt{2}}$ is in this category [$\ket{\psi}=\frac{1}{\sqrt{2}}[{\ket{000}+\ket{111}}]$].  

\subsubsection{States with four real coefficients:}
Let us examine whether there are more than two coefficients in highly entangled three-qubit states. If the coefficients are arranged accordingly, the second terms (cross terms) on the right-hand side of expressions (\ref{4},\ref{5},\ref{6}) will vanish. In this case, the condition required is that the product of any two sets of complementary coefficients is negative. For example, states with complementary coefficients $a,h,b,g$ having $ahbg<0$ will have the concurrence 1. An example of such a maximally entangled state is $\ket{\psi}=\frac{1}{\sqrt{4}}[{\ket{000}+\ket{001}-\ket{110}+\ket{111}}]$ is a maximally entangled state.

\subsubsection{States with eight real coefficients:}
There are even highly entangled states with eight coefficients. There will be four negative coefficients, three from any two sections (1st section: $a,b,c,d$ and 2nd section: $h,g,f,e$), and 4th one will be from another alternative section with the complementary element to get the zero contribution from the 2nd terms(cross terms) of the RHS of the above three expressions (\ref{4},\ref{5},\ref{6}). One example is $a,b,c$ from the 1st section, and the other is $e$ from the 2nd section, or $a$ from the 1st section and $e,f,g$ from the 2nd section. In this way the two possibilities can be $\ket{\psi}=\frac{1}{\sqrt{8}}[-\ket{000}-\ket{001}-\ket{010}+\ket{011}-\ket{100}+\ket{101}+\ket{110}+\ket{111}]$ or $\ket{\psi}=\frac{1}{\sqrt{8}}[-\ket{000}+\ket{001}+\ket{010}+\ket{011}-\ket{100}-\ket{101}-\ket{110}+\ket{111}]]$  \\

In conclusion for three-qubit pure states, we have the criterion- $$(E^2_{A}, E^2_{B}, E^2_{C} \le 1 )$$.
We have tried to get all possible cases to get the maximum concurrence value of 1 for each bipartition.
We have to remember that the odd number coefficients do not allow zero contribution for the 2nd terms of the RHS of the above three expressions (\ref{4},\ref{5},\ref{6}), and that is the reason not to have the odd number of coefficients for the maximally entangled states. We also want to add to our observation that all the states within each subsection are locally equivalent.

\subsection{Four qubit Maximally Entangled States:}
Using the concept of the three-qubit maximally entanglement criterion, one can find the concurrence for the four-qubit system. We consider a four-qubit system with qubits labelled by $A, B, C, D$. Let us also consider $\ket{\psi}$ to be the pure state with density matrix $\rho=\ket{\psi}\bra{\psi}$. Taking all the combinations, the minimum possible bipartitions for this case are-  {$A|BCD$}, {$B|ACD$}, {$C|ABD$}, $D|ABC$, $AB|CD$, $AC|BD$, $AD|BC$. 
Considering all independent bipartitions, we have the global measure of entanglement for the four-qubit system as:
$$E=E_A+E_B+E_C+E_D+E_{AB}+E_{AC}+E_{AD}$$

If we represent the general pure four qubit state (with qubits labelled by $A, B, C, D$) as
\begin{equation}\label{11}
    \begin{aligned}
        \ket{\psi}  &=  a\ket{0000}+b\ket{0001}+c\ket{0010}+d\ket{0011}+e\ket{0100}+f\ket{0101}+g\ket{0110}\\
        &+h\ket{0111}+i\ket{1000}+j\ket{1001}+k\ket{1010}+l\ket{1011}+m\ket{1100}+n\ket{1101}\\
        &+o\ket{1110}+p\ket{1111}       
    \end{aligned}
\end{equation}
One can easily check that the measurement of $E_A$, $E_B$, $E_C$, $E_D$ will be $\le 1$.\\
However, for the other bipartitions, we can get higher concurrence for the pure state with real coefficients.\\
We have derived the below conditions for $E_{AB}, E_{AC}, E_{AD}>1$:
{\footnotesize{
\begin{equation}\label{12}
    \begin{aligned}
    &|a\Bar{e}+b\Bar{f}+c\Bar{g}+d\Bar{h}|^2+|a\Bar{i}+b\Bar{j}+c\Bar{k}+d\Bar{l}|^2+|a\Bar{m}+b\Bar{n}+c\Bar{o}+d\Bar{p}|^2+|e\Bar{m}+f\Bar{n}+g\Bar{o}+h\Bar{p}|^2\\
        & +|i\Bar{m}+j\Bar{n}+k\Bar{o}+l\Bar{p}|^2+|e\Bar{i}+f\Bar{j}+g\Bar{k}+h\Bar{l}|^2<(1/8)
    \end{aligned}
\end{equation}
\begin{equation}\label{13}
\begin{aligned}
    &|a\Bar{c}+b\Bar{d}+e\Bar{g}+f\Bar{h}|^2+|a\Bar{i}+b\Bar{j}+e\Bar{m}+f\Bar{n}|^2 
    +|a\Bar{k}+b\Bar{l}+e\Bar{o}+f\Bar{p}|^2+|c\Bar{k}+d\Bar{l}+g\Bar{o}+h\Bar{p}|^2 \\
    & +|i\Bar{k}+j\Bar{l}+m\Bar{o}+n\Bar{p}|^2+|i\Bar{c}+j\Bar{d}+m\Bar{g}+n\Bar{h}|^2<(1/8)
\end{aligned}
\end{equation}
\begin{equation}\label{14}
    \begin{aligned}
        &|a\Bar{b}+c\Bar{d}+e\Bar{f}+g\Bar{h}|^2+|a\Bar{i}+c\Bar{k}+e\Bar{m}+g\Bar{o}|^2 
        +|a\Bar{j}+c\Bar{l}+e\Bar{n}+g\Bar{p}|^2+|b\Bar{j}+d\Bar{l}+f\Bar{n}+h\Bar{p}|^2\\
        & +|i\Bar{j}+k\Bar{l}+m\Bar{n}+o\Bar{p}|^2+|b\Bar{i}+d\Bar{k}+f\Bar{m}+h\Bar{o}|^2<(1/8)
    \end{aligned}
\end{equation}
}}

Computationally, we can optimize the state to a maximally entangled state. Here the idea is, as we know, the maximum concurrence for the single bipartitions ($E_A$, $E_B$, $E_C$, $E_D$) is 1, then we first try to figure out this condition for a four-qubit state. After that, we will go for the double bipartition case. With this condition, we will discuss various cases in the sections below.

\subsubsection{States with two real coefficients:}\label{3.2.1}
For all sixteen real coefficients with equal probability, one cannot archive the concurrence for an entangled state of more than 1. But one can have the same value as 1. If only the conjugate coefficient of a pure state (\ref{11}) exists, e.g. [$(a,p); (b,o); (c,n); (d,m); (e,l); (f,k); (g,j); (h,i)$], whatever the sign, we have a maximally entangled state. Similar to the three-qubit case, the four-qubit GHZ state $\ket{\psi}=\frac{1}{\sqrt{2}}[{\ket{0000}+\ket{1111}}]$ is in this category (as $a=p=\frac{1}{\sqrt{2}}$). 

\subsubsection{States with four real coefficients:}\label{3.2.2}
Like the three-qubit case, we will have the maximum concurrence when we have more than two coefficients for the subsystems. If the arrangement of the coefficients is that way, the cross terms in the trace operations of the $E_A$, $E_B$, $E_C$, and $E_D$ vanish like the case of the three-qubit state. The corresponding condition is any two sets from the sets [$(a,p); (b,o); (c,n); (d,m); (e,l); (f,k); $ $(g,j); (h,i)$] satisfying their multiplication as (-)ve. The condition for this case is that the states with conjugate coefficients $a,p,f,k$ if $apfk<0$. As an example we can represent a state $\ket{\psi}=\frac{1}{\sqrt{4}}[{\ket{0000}+\ket{0101}}-{\ket{1010}+\ket{1111}}]$ belongs in this category .

\subsubsection{States with eight real coefficients:}\label{3.2.3}
There are even highly entangled states with eight coefficients. There will be four negative coefficients, three from any two sections (1st section: $a,b,c,d,e,f,g,h$; 2nd section: $p,o,n,m,l,k,j,i$), and another will be another section with the conjugate element. The state with $a,b,c$ from 1st section and $m$ from 2nd section or $a$ from 1st section and $k,j,i$ from 2nd section has the maximum concurrence value 1. We have these two examples corresponding these two cases- $\ket{\psi}=\frac{1}{\sqrt{8}}[-{\ket{0000}-\ket{0001}}-{\ket{0010}+\ket{0011}+\ket{0100}+\ket{0101}}$ $+{\ket{0110}-\ket{1100}}]$ and $\ket{\psi}=\frac{1}{\sqrt{8}}[{-\ket{0000}+\ket{0001}}+{\ket{0010}+\ket{0011}}+\ket{0100}$ $-\ket{1000}-{\ket{1001}-\ket{1010}}]$.

\subsubsection{States with sixteen real coefficients:}\label{3.2.4}
There will be eight negative coefficients, six from any two sections $(a,b,c,$ $d,e,f,g,h; p,o,n,m,l,k,j,i)$, and another two will be another section with the conjugate element. Ex- The state having $a,b,c,d,e,f$ from 1st section and $i,j$ from 2nd section or $c,d$ from 1st section and $p,o,l,k,j,i$ from 2nd section ( Ex-$\ket{\psi}=\frac{1}{\sqrt{16}}[-\ket{0000}-\ket{0001}-\ket{0010}-\ket{0011}-\ket{0100}-\ket{0101}+\ket{0110}+\ket{0111}-\ket{1000}-\ket{1001}+\ket{1010}+\ket{1011}+\ket{1100}+\ket{1101}+\ket{1110}+\ket{1111}]$ and $\ket{\psi}=\frac{1}{\sqrt{16}}[\ket{0000}+\ket{0001}-\ket{0010}-\ket{0011}+\ket{0100}+\ket{0101}+\ket{0110}+\ket{0111}-\ket{1000}-\ket{1001}-\ket{1010}-\ket{1011}+\ket{1100}+\ket{1101}-\ket{1110}-\ket{1111}]$ ). \\

These four criteria are corresponding to $(E^2_{A}, E^2_{B}, E^2_{C}, E^2_{D} = 1 )$. We already know that for a single cut, the concurrence is 1, but for a double cut, the value will be higher than that. \\

Remember again that the odd number coefficients do not allow zero contribution for the 2nd (cross) terms of the RHS of the above three expressions (\ref{12}, \ref{13}, \ref{14}). In that case, the concurrence for the single cut will be less than 1.
In the conclusion we have $$(E^2_{A}, E^2_{B}, E^2_{C}, E^2_{D} \le 1 )$$

The subsections (\ref{3.2.1}, \ref{3.2.2}, \ref{3.2.3}, \ref{3.2.4}) say the criteria to get the maximum concurrence value 1 for a pure state for the single cut bipartition. This condition will be satisfied for all maximally four qubit pure entangled states.

\subsection{five qubit Maximally Entangled States:}

We consider a five-qubit system with qubits labeled by $A, B, C, D, E$.Let us also consider $\ket{\psi}$ be the pure state with density matrix $\rho=\ket{\psi}\bra{\psi}$. Taking all the combinations, all possible bipartitions for this case are-\\  
$A|BCDE$, $B|ACDE$, $C|ABDE$, $D|ABCE$, $E|ABCD$, $AB|CDE$, $AC|BDE$, $AD|BCE$, $AE|BCD$, $BC|ADE$, $BD|ACE$, $BE|ACD$, $CD|ABE$, $CE|ABD$, $DE|ABC$.

Considering all independent bipartitions, we have the global measure of entanglement for the five qubit system as:
$$E=E_A+E_B+E_C+E_D+E_E+E_{AB}+E_{AC}+E_{AD}+E_{AE}+E_{BC}+E_{BD}+E_{BE}$$$$+E_{CD}+E_{CE}+E_{DE}$$

Where $$E^2_{X}=2[1-tr({\rho_X}^2)]$$
A general 5 qubit pure state can be expressed as \begin{equation}\label{5-qubit}
    \begin{aligned}
        & \ket{\psi}=\\
        &[a\ket{00000}+b\ket{00001}+c\ket{00010}+d\ket{00011}+e\ket{00100}+f\ket{00101}\\
        & +g\ket{00110}+h\ket{00111}+i\ket{01000}+j\ket{01001}+k\ket{01010}+l\ket{01011}\\
        &  +m\ket{01100}+n\ket{01101}+o\ket{01110}+p\ket{01111}+q\ket{10000}+r\ket{10001}\\
        & +s\ket{10010}+t\ket{10011}+u\ket{10100}+v\ket{10101}+w\ket{10110}+x\ket{10111}\\
        & +y\ket{11000}+z\ket{11001}+\alpha\ket{11010}+\beta\ket{11011}+\gamma\ket{11100}+\delta\ket{11101}\\
        &
        +\theta\ket{11110}+\phi\ket{11111}]
    \end{aligned}
\end{equation}
Similarly in the measurement of $E_A$, $E_B$, $E_C$, $E_D$, $E_E$ we will have $\le 1$.\\
However, for the higher bipartitions or cuts, we can get higher entanglement values for the pure state with real coefficients.\\
%\newpage
The conditions for $E_{AB}, E_{AC}, E_{AD}, E_{AE}, E_{BC}, E_{BD}, E_{BE}, E_{CE}, E_{DE}>1$ have been derived and shown in the Appendix section. Like the three and four-qubit pure states, one can also examine the maximal criteria for five-qubit pure states similarly.

\begin{figure}[ht]
    \centering
    		\begin{subfigure}{0.3\textwidth}
			\begin{tikzpicture}[scale = 1.25]
				\draw[fill] (0, 0) circle [radius = 2pt];
				\node[above left] at (0, 0) {$3$};
				\draw[fill] (2.5, 0) circle [radius = 2pt];
				\node[above right] at (2.5, 0) {$2$};
				\draw[fill] (2.5, 2.2) circle [radius = 2pt];
				\node[above right] at (2.5, 2.2) {$1$};
				\draw [blue] (2.5, 2.2)--(2.5, 0);
				\draw [red] (2.5, 2.2)--(0, 0);
				\draw [green] (2.5, 0)--(0, 0);
			\end{tikzpicture}
			\caption{Three qubit symmetric graph state}
			\label{Fig11}
		\end{subfigure}
		~
		\begin{subfigure}{0.3\textwidth}
			\begin{tikzpicture}[scale = 1.0]
				\draw[fill] (0, 0) circle [radius = 2pt];
				\node[above right] at (0, 0) {$3$};
				\draw[fill] (2, 0) circle [radius = 2pt];
				\node[above left] at (2, 0) {$2$};
				\draw[fill] (2, 2) circle [radius = 2pt];
				\node[below left] at (2, 2) {$1$};
				\draw [blue] (-0.45,-0.3)--(2.3,-0.3)..controls (2.67,-0.09)..(2.5,0.65)--(2.25,2.4)..controls (2.15,2.6)..(2.,2.5)--(-0.5,0)..controls (-0.56,-0.2)..(-0.45,-0.3);
				
				%\node[below] at (1,-1) {$Fig-1$};
			\end{tikzpicture}
			\caption{Three qubit hypergraph State}
			\label{Fig12}
		\end{subfigure}
  \begin{subfigure}{0.3\textwidth}
\centering
\begin{tikzpicture}[scale = 0.925]
		\node[above right] at (0, 0) {$3$};
  				\draw[fill] (0, 0) circle [radius = 2pt];
\draw[fill] (0, 3) circle [radius = 2pt];
				\node[below right] at (0, 3) {$1$};
				
				\draw[fill] (3, 3) circle [radius = 2pt];
				\node[below left] at (3, 3) {$2$};
    \draw[blue] (0, 0)--(0, 3);
        \draw[magenta] (3, 3)--(0, 3);

\end{tikzpicture}
\caption{Three qubit GHZ state}
\vspace{0.45cm}
\label{Fig13}
\end{subfigure}
    \caption{Three qubit states}
    \label{Fig1}
\end{figure}
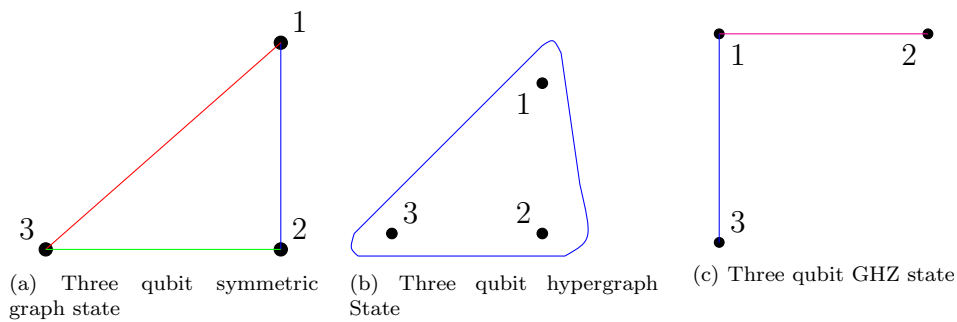

\section{More on AME States:} \label{sec3}
It is known that all reduced density matrices $\rho$ corresponding to $k$ qudits ($d$ dimension) can be represented as\cite{11s} 
\begin{equation}
    \rho=\frac{1}{d^k}I_{d^k}; \forall{k}\leq\frac{N}{2}
\end{equation} 
For a maximally entangled subsystem with a single-cut bipartition, the value of $k$ will be 1. Similarly, $k$=2 represents the double-qubit partition, and so on. Here, we have focused on the qubit system where $d$=2. Determining the entanglement corresponding to the $N$ qubit states is straightforward using $k$-uniformity. However, it is still not clear how many AME states exist in this world. Next, we will examine the maximal concurrence criterion and its applicability to AME states. Before that, we established a relationship between the number $k$ and maximum concurrence. The generalized form for the value of maximum concurrence corresponding to the $k$ number of a qubit system has been derived as 
\begin{equation}\label{21}
    E|_{max}=\sqrt{\frac{2^k-1}{2^{k-1}}}
\end{equation} 
The concurrence $E$ will never reach 2, but with the increase of the $k$ value, this will try to reach closer to 2.

\subsection{Three qubit Absolutely Maximally Entangled (AME) States:}
We consider a three qubit pure state having three bipartitions $A|BC$, $B|AC$, $C|AB$ when the qubit levels are $A, B, C$. For the bipartition $A|BC$, we have the von Neumann entropy being an AME(3,2) state as
$$S(\rho_A)=Tr[\rho_A{log}_d\rho_A]=-\sum_i\lambda_ilog_2\lambda_i= \lfloor 3/2 \rfloor = 1 $$
From the measurement of concurrence
$$E_A^2=2[1-tr({\rho_A}^2)]=2[1-(\lambda_i^2)]=1$$
Combining these two criteria we have $$2[1-(\lambda_1^2+\lambda_2^2)]=-\lambda_1log_2\lambda_1-\lambda_2log_2\lambda_2=1$$
From here $$\lambda_1=\lambda_2=0.5$$

For the three-qubit pure state, the maximum total bipartite concurrence is 3, having maximally mixed with each bipartition, which confirms that the concurrence for the three-qubit AME(3,2) will be 3. The three-qubit GHZ state belongs to this category of AME(3,2). The three-qubit GHZ state (\ref{Fig13}) in the graphical format can be stated as:
$$\ket{\psi}=\frac{1}{\sqrt{8}}[\ket{000}+\ket{001}+\ket{010}+\ket{011}+\ket{100}-\ket{101}-\ket{110}-\ket{111}]$$
The concurrence for each bipartition will be $E_A=E_B=E_C=1$.  \\
A three-qubit symmetric graph state (\ref{Fig11}) has been studied for the purpose of metrological uses\cite{bag2022achieving} showing the maximum concurrence for each bipartition. The state can be expressed as:
$$\ket{\psi}=\frac{1}{\sqrt{8}}[\ket{000}+\ket{001}+\ket{010}-\ket{011}+\ket{100}-\ket{101}-\ket{110}-\ket{111}]$$
Again, the concurrence for each bipartition will be $E_A=E_B=E_C=1$.\\
It confirms that these two states discussed above are AME states. One can conclude that all the three qubit maximally entangled states having concurrence 1 for each bipartition are Absolutely Maximally Entangled states.

\begin{figure}[ht]
    \centering
    \begin{subfigure}{0.4\textwidth}
	\begin{tikzpicture}[scale = 0.85]
		\draw[fill] (0, 0) circle [radius = 2pt];
				\node[above right] at (0, 0) {$1$};
				\draw[fill] (0, 3) circle [radius = 2pt];
				\node[below right] at (0, 3) {$2$};
				\draw[fill] (3, 0) circle [radius = 2pt];
				\node[above left] at (3, 0) {$4$};
				\draw[fill] (3, 3) circle [radius = 2pt];
				\node[below left] at (3, 3) {$3$};
				\draw [red](-.5, 1.5) .. controls (-.5, -.5) .. (1.5, -.5) .. controls (4, -.5) .. (1.75, 1.75) .. controls (-.5, 4) .. (-.5, 1.5);
				\draw[red] (1.5, 2.1) .. controls (-1.25, -.75) .. (1.5, -.75) .. controls (3.75, -.75) .. (3.5, 2) .. controls (3.4, 4) .. (1.5, 2.1);
				\draw[blue] (1.5, 3.75) .. controls (-1, 3.75) .. (-1, 1.5) .. controls (-1, -1.5) .. (2, 1.5) .. controls (4.25, 3.75) .. (1.5, 3.75);
				\draw[brown] (1.5, 4) .. controls (-1.75, 4) .. (1, 1.25) .. controls (4, -1.5) .. (4, 1.5) .. controls (4, 4) .. (1.5, 4);
\end{tikzpicture}
\caption{Four qubit three uniform hypergraph state}
\label{Fig21}
\end{subfigure}
		~
\begin{subfigure}{0.4\textwidth}
\centering
\begin{tikzpicture}[scale = 1.3]
		\node[above right] at (0, 0) {$4$};
  				\draw[fill] (0, 0) circle [radius = 2pt];
\draw[fill] (0, 3) circle [radius = 2pt];
				\node[below right] at (0, 3) {$1$};
				\draw[fill] (3, 0) circle [radius = 2pt];
				\node[above left] at (3, 0) {$3$};
				\draw[fill] (3, 3) circle [radius = 2pt];
				\node[below left] at (3, 3) {$2$};
    \draw[blue] (0, 0)--(0, 3);
        \draw[green] (3, 0)--(0, 3);
        \draw[magenta] (3, 3)--(0, 3);

\end{tikzpicture}
\vspace{0.35cm}
\caption{Four qubit GHZ state}
\vspace{0.45cm}
\label{Fig22}
\end{subfigure}
    \caption{Four qubit states}
    \label{Fig2}
\end{figure}
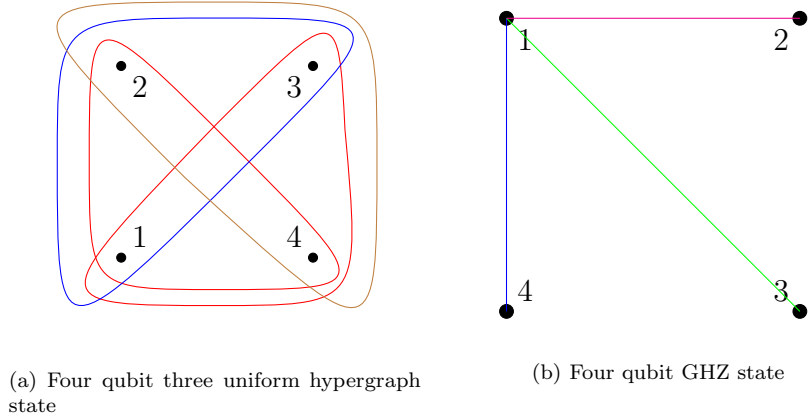

\subsection{Criterion for four qubit Absolutely Maximally Entangled (AME) States:}
As for the single bipartition, we cannot get the concurrence beyond 1. In this context, for the four qubit AME states corresponding $\lambda{s}$, the eigenvalues will be $\frac{1}{2}$ for the reduced density matrix of single qubit bipartition. In that way, the definition of the AME state has been satisfied. For the double bipartition in general the von Neumann entropy can be expressed as 
$$S(\rho)=\lfloor n/d \rfloor$$ 
but for single bipartition $$S(\rho)=1$$ 

It has been established that there is no existence of a 4-qubit AME state \cite{3s} as the required constraints are not satisfied to achieve the maximally mixed criterion for double-qubit bipartitions. According to \cite{3s}, only two of the double-qubit bipartitions are maximally mixed, while one is not. Let us explore the scenario where all three bipartitions are maximally mixed. We will examine this in the context of the four-qubit maximally entangled condition, specifically for `states with sixteen real coefficients' as discussed in Section \ref{3.2.4}.\\
Suppose a state consists of negative $a,b,c,d,e,f, i,j$ coefficients, where the single qubit concurrences corresponding to the states are 1, showing the maximal mixedness.
If we measure the entropy of double qubit bipartitions corresponding to $AB/CD, AC/BD, AD/BC$, we have $S(\rho_{AB})=S(\rho_{AC})=S(\rho_{AD})=1$ that does not imply the maximally mixed criterion. Also, the value of corresponding bipartite entanglement (concurrence) is 1, which is not maximum (according to equation \ref{21}).  

This concludes that if the single-cut bipartition satisfies the maximally mixed criterion, the double-qubit or double-cut bipartition may or may not be maximally mixed. Our focus is to get all the bipartitions to be maximally mixed for an AME state.\\

We can show that the $k$-uniformity criterion for four qubit states violates the concurrence value corresponding to a bipartition. For a double-cut bipartition when $k=2$, the corresponding concurrence will be: $\sqrt{3/2}$ using equation (\ref{21}). If a four-qubit AME state exists, then this concurrence value will be for these bipartitions $\rho_{AB}$, $\rho_{AC}$, $\rho_{AD}$. Suppose the concurrence value is $\sqrt{3/2}$ for these bipartitions. Then, the coefficient criteria (when all 16 coefficients exist) will be:
{\footnotesize{
\begin{equation}\label{22}
    \begin{aligned}
        &|a\Bar{e}+b\Bar{f}+c\Bar{g}+d\Bar{h}|^2+|a\Bar{i}+b\Bar{j}+c\Bar{k}+d\Bar{l}|^2
        +|a\Bar{m}+b\Bar{n}+c\Bar{o}+d\Bar{p}|^2+|e\Bar{m}+f\Bar{n}+g\Bar{o}+h\Bar{p}|^2\\
        & +|i\Bar{m}+j\Bar{n}+k\Bar{o}+l\Bar{p}|^2+|e\Bar{i}+f\Bar{j}+g\Bar{k}+h\Bar{l}|^2=0
    \end{aligned}
\end{equation}
\begin{equation}\label{23}
\begin{aligned}
    &|a\Bar{c}+b\Bar{d}+e\Bar{g}+f\Bar{h}|^2+|a\Bar{i}+b\Bar{j}+e\Bar{m}+f\Bar{n}|^2
    +|a\Bar{k}+b\Bar{l}+e\Bar{o}+f\Bar{p}|^2+|c\Bar{k}+d\Bar{l}+g\Bar{o}+h\Bar{p}|^2 \\
    & +|i\Bar{k}+j\Bar{l}+m\Bar{o}+n\Bar{p}|^2+|i\Bar{c}+j\Bar{d}+m\Bar{g}+n\Bar{h}|^2=0
\end{aligned}
\end{equation}
\begin{equation}\label{24}
    \begin{aligned}
        &|a\Bar{b}+c\Bar{d}+e\Bar{f}+g\Bar{h}|^2+|a\Bar{i}+c\Bar{k}+e\Bar{m}+g\Bar{o}|^2
        +|a\Bar{j}+c\Bar{l}+e\Bar{n}+g\Bar{p}|^2+|b\Bar{j}+d\Bar{l}+f\Bar{n}+h\Bar{p}|^2\\
        & +|i\Bar{j}+k\Bar{l}+m\Bar{n}+o\Bar{p}|^2+|b\Bar{i}+d\Bar{k}+f\Bar{m}+h\Bar{o}|^2=0
    \end{aligned}
\end{equation}
}}
If one can use the criterion of the section- \ref{3.2.4} having $a,b,c,d,e,f, i,j$ are negative, then the above conditions in equations (\ref{22}), (\ref{23}) and (\ref{24}) will not be satisfied together. In conclusion, this example proves the non-existence of a four qubit AME state having maximally mixed single bit bipartitions.

Suppose we focus only on the double-qubit bipartitions. In this case, the pure state would not be feasible in a real scenario, as there is no possibility \cite{3s} of obtaining four eigenvalues of $1/4$ for each of the three bipartite reduced density matrices $\rho_{AB}$, $\rho_{AC}$, $\rho_{AD}$. Let us check the criterion for the case when there will be a four-qubit AME state. Consider a hypothetical case where we have a four-qubit AME state with the corresponding equal eigenvalues as $1/4$ of the three double-qubit bipartitions showing $S(\rho_{AB})=S(\rho_{AC})=S(\rho_{AD})=2$ i.e. maximally mixed. ( As $\lambda_1=\lambda_2=\lambda_3=\lambda_4=\lambda=(1/4)$ $\implies$ $S(\rho_{AB})=S(\rho_{AC})=S(\rho_{AD})=4\times(-\lambda{log_2\lambda})=2$.)
 The corresponding bipartite concurrences will be $\sqrt{3/2}$. ($E_{AB}=E_{AC}=E_{AD}=\sqrt{2[1-{(\lambda_1^2+\lambda_2^2+\lambda_3^2+\lambda_4^2)}]}=\sqrt{2[1-4\lambda^2]}=\sqrt{3/2}$). This condition will never be satisfied by maximal criterion for the single bipartition($E_A=E_B=E_C=1$ or $S(\rho_A)=S(\rho_B)=S(\rho_C)=1$). We will never get this case, which is already explained \cite{3s}. Even Sudbery and Higuchi \cite{3s} showed a different angle for a special four qubit state ($\psi_4=\frac{1}{2}[\ket{0000}+\ket{0111}+\ket{1001}+\ket{1110}]$). Not all double-cut bipartitions have the highest entropy, i.e., two. If we measure the concurrence corresponding to the double-cut bipartitions having entropy two have the maximum concurrence $\sqrt{3/2}$ but we do not have the maximum concurrence and entropy for the 3rd bipartition. This ends up with the conclusion that there is no existence of four qubit AME states.

\begin{figure}[ht]
    \centering
            \begin{subfigure}{0.4\textwidth}
			\centering
			\begin{tikzpicture}[scale = 1.4]
		\draw[fill] (-1.5, 3) circle [radius = 2pt];
\node[above left] at (-1.5, 3) {$5$};
\draw[fill] (-0.75, 1.5) circle [radius = 2pt];
		\node[below left] at (-0.75, 1.5) {$4$};
\draw[fill] (0.75, 1.5) circle [radius = 2pt];
\node[below right] at (0.75, 1.5) {$3$};
 \draw[fill] (1.5, 3) circle [radius = 2pt];
\node[above right] at (1.5, 3) {$2$};
\draw[fill] (0,4) circle [radius = 2pt];
\node[above] at (0, 4) {$1$};
	\draw [blue] (0, 4)--(-1.5, 3);
	\draw [red] (0, 4)--(1.5, 3);
	\draw [green] (1.5, 3)--(0.75, 1.5);
 	\draw [magenta] (0.75, 1.5)--(-0.75, 1.5);
 	\draw [teal] (-1.5, 3)--(-0.75, 1.5);

\end{tikzpicture}
			\caption{Five qubit two uniform graph state}
			\label{Fig31}
\end{subfigure}
		~
            \begin{subfigure}{0.4\textwidth}
	\begin{tikzpicture}[scale = 1.1]
		\draw[fill] (0, 0) circle [radius = 2pt];
		\node[above left] at (0, 0) {$5$};
		\draw[fill] (1.5, -0.5) circle [radius = 2pt];
		\node[above left] at (1.5, -0.5) {$4$};
\draw[fill] (3, -0.5) circle [radius = 2pt];
\node[above left] at (3, -0.5) {$3$};
 \draw[fill] (4.5, 0) circle [radius = 2pt];
		\node[below right] at (4.5, 0) {$2$};
	\draw [blue] (2.5, 2.2)--(1.5, -0.5);
	\draw [red] (2.5, 2.2)--(0, 0);
	\draw [green] (2.5, 2.2)--(3, -0.5);
 	\draw [magenta] (2.5, 2.2)--(4.5, 0);
\draw[fill] (2.5, 2.2) circle [radius = 2pt];
\node[above] at (2.5, 2.5) {$1$};
\end{tikzpicture}
			\caption{Five qubit GHZ state}
			\label{Fig32}
		\end{subfigure}
    \caption{Five qubit states}
    \label{Fig3}
\end{figure}
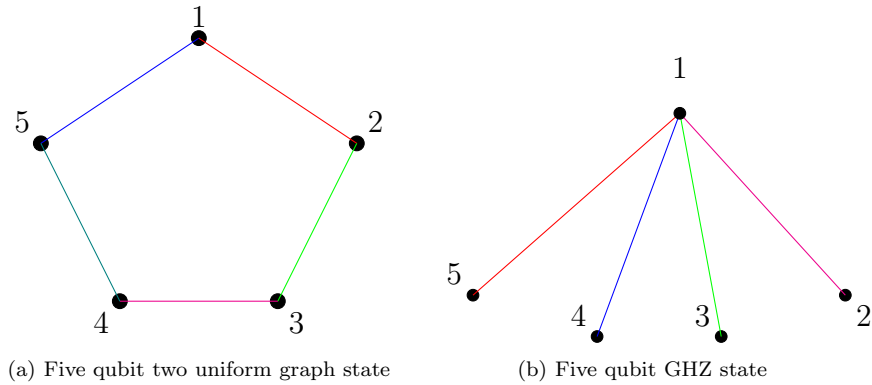

\subsection{Five qubit Absolutely Maximally Entangled (AME) States:}
It is confirmed that if AME states exist, these will always have the maximum concurrence value. In this section we will show how the existing five qubit AME state satisfy our maximal concurrence criterion. We consider a five qubit AME state (\ref{Fig31}) as:
$$\ket{\psi}={\frac{1}{\sqrt{32}}}$$$$[\ket{00000}+\ket{00001}+\ket{00010}-\ket{00011}+\ket{00100}+\ket{00101}-\ket{00110}+\ket{00111}$$
$$+\ket{01000}+\ket{01001}+\ket{01010}-\ket{01011}-\ket{01100}-\ket{01101}+\ket{01110}-\ket{01111}$$
$$+\ket{10000}-\ket{10001}+\ket{10010}+\ket{10011}+\ket{10100}-\ket{10101}-\ket{10110}-\ket{10111}$$
$$-\ket{11000}+\ket{11001}-\ket{11010}-\ket{11011}+\ket{11100}-\ket{11101}-\ket{11110}-\ket{11111}]$$
The concurrence for each bipartition will be $E_A=E_B=E_C=E_D=E_E=1; %0.9997364; 
E_{AB}=E_{AC}=E_{AD}=E_{AE}=E_{BC}=E_{BD}=E_{BE}=E_{CD}=E_{CE}=E_{DE}=1.224$. These values of concurrence satisfy the equation (\ref{21}). This will also be satisfied if we check the maximum concurrence criterion for this state (Appendix \ref{A}). Hence, it proves that all AME states have the maximum concurrence for each bipartition. Even in fact, the generalized concurrence criterion has been followed by this five-qubit AME state.

\section{Criterion for EME states:} \label{sec4}
As it has been proved that the maximum concurrence for the single bipartition of any qubit state is 1, then the EME states have a concurrence of 1 for any bipartition. For three-qubit maximally entangled states having concurrence, 1 is the EME state. Still, for the $n>3$ case, all the maximally entangled states are not the EME states, as for the cases of double bipartition, one can have the concurrence $>1$.\\
In another way, there exist a few states that are under the EE(equal entanglement) class, but there may not be a maximum. Ex- The concurrences of the three-qubit hypergraph state for each partition is not one (maximum) but show EE nature.
The only exists three qubit hypergraph state (\ref{Fig12}) can be stated as 
$$\ket{\psi}=\frac{1}{\sqrt{8}}[\ket{000}+\ket{001}+\ket{010}+\ket{011}+\ket{100}+\ket{101}+\ket{110}-\ket{111}]$$
The concurrence for each bipartition will be $E_A=E_B=E_C=0.866$ which is < 1 implies not maximum.\\
For the double bipartition case, we have the $4\times{4}$ reduced density matrix and four eigenvalues ($\lambda{s}$) satisfy $\lambda_1^2+\lambda_2^2+\lambda_3^2+\lambda_4^2=\frac{1}{2}$ using the criterion of maximum concurrence and the maximum concurrence of a four qubit state for each partition will be 1. Ex- A graph state having this form. A four qubit EME state in \ref{Fig21} is represented as
$$\ket{\psi}={\frac{1}{\sqrt{16}}}[\ket{0000}+\ket{0001}+\ket{0010}+\ket{0011}+\ket{0100}+\ket{0101}+\ket{0110}-\ket{0111}$$
$$+\ket{1000}+\ket{1001}+\ket{1010}-\ket{1011}+\ket{1100}-\ket{1101}-\ket{1110}+\ket{1111}]$$
If we measure the concurrence for each bipartition, it will be $E_A=E_B=E_C=E_D=E_{AB}=E_{AC}=E_{AD}=1$.\\
Even this is the same for a four qubit GHZ state (\ref{Fig22}) if we represent this state in graph form
$$\ket{\psi}={\frac{1}{\sqrt{16}}}[\ket{0000}+\ket{0001}+\ket{0010}+\ket{0011}+\ket{0100}+\ket{0101}+\ket{0110}+\ket{0111}$$
$$+\ket{1000}-\ket{1001}-\ket{1010}+\ket{1011}-\ket{1100}+\ket{1101}+\ket{1110}-\ket{1111}]$$

The same trend follows for the five qubit EME states. The graph for the five qubit GHZ state (\ref{Fig32}) is in this category, having the maximum concurrence. The five qubit EME(GHZ) state:
$$\ket{\psi}={\frac{1}{\sqrt{32}}}$$$$[\ket{00000}+\ket{00001}+\ket{00010}-\ket{00011}+\ket{00100}-\ket{00101}-\ket{00110}-\ket{00111}$$
$$+\ket{01000}-\ket{01001}-\ket{01010}-\ket{01011}-\ket{01100}-\ket{01101}-\ket{01110}+\ket{01111}$$
$$+\ket{10000}-\ket{10001}-\ket{10010}-\ket{10011}-\ket{10100}-\ket{10101}-\ket{10110}+\ket{10111}$$
$$-\ket{11000}-\ket{11001}-\ket{11010}+\ket{11011}-\ket{11100}+\ket{11101}+\ket{11110}+\ket{11111}]$$
The concurrence for each bipartition will be $E_A=E_B=E_C=E_D=E_E=E_{AB}=E_{AC}=E_{AD}=E_{AE}=E_{BC}=E_{BD}=E_{BE}=E_{CD}=E_{CE}=E_{DE}=1$. One can also agree with the statement that all GHZ states are the EME states, having equal concurrence for each partition.

\section{Discussions and Conclusions:} \label{sec5}
Concurrence is a robust measurement for quantifying entanglement. Our exploration centered on establishing a connection between the criterion for maximum concurrence and pure states that are absolutely maximally entangled (AME). The maximum concurrence criterion is based on the measurement of concurrence for a bipartition where the trace of the square of the corresponding reduced density matrix is minimum. 

We again demonstrated that pure AME qubit states achieve the highest concurrence across all possible bipartitions. Additionally, we revealed that pure states characterized by an odd number of coefficients fail to meet the AME criterion. Specifically, the W class of states never exhibits AME behaviour.

This criterion gives the information that, as we cannot have a concurrence of more than 1 for any single qubit bipartition, it helps to find out more AME states having the von Neumann entropy as 1. The maximally entangled states having concurrence 1 for each bipartition are EME (Equal Maximally Entangled) states for the three-qubit cases. This implies that all three-qubit AME states are equal to EME states, but this is not true for the higher ($n>3$) qubit pure states. For the higher qubit pure state, we strongly say that the EME states will be the class of GHZ-like states where the entanglement, in this case for each bipartition, will be 1.

In this paper, using the maximum concurrence criterion, we have thoroughly studied all possibilities of having pure ME (Maximally Entangled), AME (Absolutely Maximally Entangled), and EME (Equal Maximally Entangled) states in the context of three and higher qubits cases. However, we limit our studies to qubit constituents because the complexity increases with the higher dimensional states. In the future, one can explore the higher-dimensional quantum states to investigate the maximum concurrence criterion, and this complexity can be easily handled using computational approaches.

\bibliographystyle{elsarticle-num} 
\bibliography{library}

\begin{thebibliography}{10}
\expandafter\ifx\csname url\endcsname\relax
  \def\url#1{\texttt{#1}}\fi
\expandafter\ifx\csname urlprefix\endcsname\relax\def\urlprefix{URL }\fi
\expandafter\ifx\csname href\endcsname\relax
  \def\href#1#2{#2} \def\path#1{#1}\fi

\bibitem{walter2013entanglement}
M.~Walter, B.~Doran, D.~Gross, M.~Christandl, Entanglement polytopes: multiparticle entanglement from single-particle information, Science 340~(6137) (2013) 1205--1208.

\bibitem{szalay2015multipartite}
S.~Szalay, Multipartite entanglement measures, Physical Review A 92~(4) (2015) 042329.

\bibitem{PhysRevA.72.062108}
W.-X. Yang, Z.-M. Zhan, J.-H. Li, Efficient scheme for multipartite entanglement and quantum information processing with trapped ions, Physical Review A—Atomic, Molecular, and Optical Physics 72~(6) (2005) 062108.

\bibitem{PhysRevLett.126.080502}
Z.~Ren, W.~Li, A.~Smerzi, M.~Gessner, Metrological detection of multipartite entanglement from young diagrams, Physical Review Letters 126~(8) (2021) 080502.

\bibitem{campbell2024series}
E.~Campbell, A series of fast-paced advances in quantum error correction, Nature Reviews Physics 6~(3) (2024) 160--161.

\bibitem{1s}
P.~Facchi, G.~Florio, G.~Parisi, S.~Pascazio, Maximally multipartite entangled states, Physical review A 77~(6) (2008) 060304.

\bibitem{2s}
H.~B.-P. N~Gisin, Bell inequality, bell states and maximally entangled states for n qubits, Physics Letters A 246~(1–2) (1998) 1--6.

\bibitem{3s}
A.~Higuchi, A.~Sudbery, How entangled can two couples get?, Physics Letters A 273~(4) (2000) 213--217.

\bibitem{4s}
A.~J. Scott, Multipartite entanglement, quantum-error-correcting codes, and entangling power of quantum evolutions, Physical Review A—Atomic, Molecular, and Optical Physics 69~(5) (2004) 052330.

\bibitem{5s}
I.~D. Brown, S.~Stepney, A.~Sudbery, S.~L. Braunstein, Searching for highly entangled multi-qubit states, Journal of Physics A: Mathematical and General 38~(5) (2005) 1119.

\bibitem{6s}
A.~Borras, A.~Plastino, J.~Batle, C.~Zander, M.~Casas, A.~Plastino, Multiqubit systems: highly entangled states and entanglement distribution, Journal of Physics A: Mathematical and Theoretical 40~(44) (2007) 13407.

\bibitem{8s}
P.~Facchi, G.~Florio, U.~Marzolino, G.~Parisi, S.~Pascazio, Classical statistical mechanics approach to multipartite entanglement, Journal of Physics A: Mathematical and Theoretical 43~(22) (2010) 225303.

\bibitem{9s}
G.~Gour, N.~R. Wallach, All maximally entangled four-qubit states, Journal of Mathematical Physics 51~(11) (2010).

\bibitem{10s}
X.-w. Zha, H.-y. Song, J.-x. Qi, D.~Wang, Q.~Lan, A maximally entangled seven-qubit state, Journal of Physics A: Mathematical and Theoretical 45~(25) (2012) 255302.

\bibitem{11s}
W.~Helwig, W.~Cui, J.~I. Latorre, A.~Riera, H.-K. Lo, Absolute maximal entanglement and quantum secret sharing, Physical Review A—Atomic, Molecular, and Optical Physics 86~(5) (2012) 052335.

\bibitem{12s}
L.~Arnaud, N.~J. Cerf, Exploring pure quantum states with maximally mixed reductions, Physical Review A—Atomic, Molecular, and Optical Physics 87~(1) (2013) 012319.

\bibitem{13s}
C.~Kl{\"o}ckl, M.~Huber, Characterizing multipartite entanglement without shared reference frames, Physical Review A 91~(4) (2015) 042339.

\bibitem{14s}
D.~Goyeneche, D.~Alsina, J.~I. Latorre, A.~Riera, K.~{\.Z}yczkowski, Absolutely maximally entangled states, combinatorial designs, and multiunitary matrices, Physical Review A 92~(3) (2015) 032316.

\bibitem{15s}
M.~Enr{\'\i}quez, I.~Wintrowicz, K.~{\.Z}yczkowski, Maximally entangled multipartite states: a brief survey, in: Journal of Physics: Conference Series, Vol. 698, IOP Publishing, 2016, p. 012003.

\bibitem{16s}
L.~Chen, D.~Zhou, Graph states of prime-power dimension from generalized cnot quantum circuit, Scientific Reports 6~(1) (2016) 27135.

\bibitem{17s}
A.~Bernal~Serrano, On the existence of absolutely maximally entangled states of minimal support, Quantum Physics Letters, 2017, vol. 6, num. 1, p. 1-3 (2017).

\bibitem{18s}
W.~Helwig, W.~Cui, Absolutely maximally entangled states: existence and applications, arXiv preprint arXiv:1306.2536 (2013).

\bibitem{19s}
W.~Helwig, Absolutely maximally entangled qudit graph states, arXiv preprint arXiv:1306.2879 (2013).

\bibitem{22s}
F.~Huber, O.~G{\"u}hne, J.~Siewert, Absolutely maximally entangled states of seven qubits do not exist, Physical review letters 118~(20) (2017) 200502.

\bibitem{32s}
F.~Huber, C.~Eltschka, J.~Siewert, O.~G{\"u}hne, Bounds on absolutely maximally entangled states from shadow inequalities, and the quantum macwilliams identity, Journal of Physics A: Mathematical and Theoretical 51~(17) (2018) 175301.

\bibitem{21s}
M.~Hein, J.~Eisert, H.~J. Briegel, Multiparty entanglement in graph states, Physical Review A—Atomic, Molecular, and Optical Physics 69~(6) (2004) 062311.

\bibitem{23s}
M.~Grassl, T.~Beth, M.~Roetteler, On optimal quantum codes, International Journal of Quantum Information 2~(01) (2004) 55--64.

\bibitem{25s}
D.~Goyeneche, Z.~Raissi, S.~Di~Martino, K.~{\.Z}yczkowski, Entanglement and quantum combinatorial designs, Physical Review A 97~(6) (2018) 062326.

\bibitem{hein2004multiparty}
M.~Hein, J.~Eisert, H.~J. Briegel, Multiparty entanglement in graph states, Physical Review A 69~(6) (2004) 062311.

\bibitem{briegel2009measurement}
H.~J. Briegel, D.~E. Browne, W.~D{\"u}r, R.~Raussendorf, M.~Van~den Nest, Measurement-based quantum computation, Nature Physics 5~(1) (2009) 19--26.

\bibitem{sarkar2021geometry}
R.~Sarkar, S.~Banerjee, S.~Bag, P.~K. Panigrahi, Geometry of distributive multiparty entanglement in 4- qubit hypergraph states, IET Quantum Communication (2021).

\bibitem{rossi2013quantum}
M.~Rossi, M.~Huber, D.~Bru{\ss}, C.~Macchiavello, Quantum hypergraph states, New Journal of Physics 15~(11) (2013) 113022.

\bibitem{dutta2019permutation}
S.~Dutta, R.~Sarkar, P.~K. Panigrahi, Permutation symmetric hypergraph states and multipartite quantum entanglement, International Journal of Theoretical Physics 58~(11) (2019) 3927--3944.

\bibitem{schlingemann2001quantum}
D.~Schlingemann, R.~F. Werner, Quantum error-correcting codes associated with graphs, Physical Review A 65~(1) (2001) 012308.

\bibitem{wagner2018analysis}
T.~Wagner, H.~Kampermann, D.~Bru{\ss}, Analysis of quantum error correction with symmetric hypergraph states, Journal of Physics A: Mathematical and Theoretical 51~(12) (2018) 125302.

\bibitem{banerjee2020quantum}
S.~Banerjee, A.~Mukherjee, P.~K. Panigrahi, Quantum blockchain using weighted hypergraph states, Physical Review Research 2~(1) (2020) 013322.

\bibitem{yang2021representations}
Y.~Yang, H.~Cao, Representations of hypergraph states with neural networks, Communications in Theoretical Physics 73~(10) (2021) 105103.

\bibitem{sarkar2021phase}
R.~Sarkar, S.~Dutta, S.~Banerjee, P.~K. Panigrahi, Phase squeezing of quantum hypergraph states, Journal of Physics B: Atomic, Molecular and Optical Physics (2021).

\bibitem{wootters1998entanglement}
W.~K. Wootters, Entanglement of formation of an arbitrary state of two qubits, Physical Review Letters 80~(10) (1998) 2245.

\bibitem{PhysRevLett.78.5022}
S.~A. Hill, W.~K. Wootters, Entanglement of a pair of quantum bits, Physical Review Letters 78 (1997) 5022--5025.

\bibitem{bhaskara2017generalized}
V.~S. Bhaskara, P.~K. Panigrahi, Generalized concurrence measure for faithful quantification of multiparticle pure state entanglement using lagrange’s identity and wedge product, Quantum Information Processing 16~(5) (2017) 118.

\bibitem{swain2022generalized}
S.~N. Swain, V.~S. Bhaskara, P.~K. Panigrahi, Generalized entanglement measure for continuous-variable systems, Physical Review A 105~(5) (2022) 052441.

\bibitem{raggio1995properties}
G.~A. Raggio, Properties of q-entropies, Journal of Mathematical Physics 36~(9) (1995) 4785--4791.

\bibitem{simon2020entropy}
S.~Morelli, K.~Claude, E.~Christopher, S.~Jens, H.~Marcus, Dimensionally sharp inequalities for the linear entropy, Linear Algebra and its Applications 584 (2020) 294--325.

\bibitem{mishra2024geometric}
A.~Mishra, S.~Mahanti, A.~K. Roy, P.~K. Panigrahi, Geometric genuine multipartite entanglement for four-qubit systems, Physics Open (2024) 100230.

\bibitem{bag2022achieving}
S.~Bag, R.~Sarkar, P.~K. Panigrahi, Achieving heisenberg limit in the phase measurement through three-qubit graph states, arXiv preprint arXiv:2208.07772 (2022).

\end{thebibliography}

\section*{Appendix-A} \label{A}
Mathematical criterion to check for maximum concurrence for each double bipartition for a five-qubit generalized pure state (\ref{5-qubit}) have been discussed here. There are 10 double bipartitions, and all conditions are given below. \\
$E_{AB}:$
$$
\begin{aligned}
    & |a\Bar{i}+b\Bar{j}+c\Bar{k}+dl+e\Bar{m}+f\Bar{n}+g\Bar{o}+h\Bar{p}|^2+|a\Bar{q}+b\Bar{r}+c\Bar{s}+d\Bar{t}+e\Bar{u}+f\Bar{v}+g\Bar{w}+h\Bar{x}|^2+\\
    & |a\Bar{y}+b\Bar{z}+c\Bar{\alpha}+d\Bar{\beta}+e\Bar{\gamma}+f\Bar{\delta}+g\Bar{\theta}+h\Bar{\phi}|^2+|i\Bar{y}+j\Bar{z}+k\Bar{\alpha}+l\Bar{\beta}+m\Bar{\gamma}+n\Bar{\delta}+o\Bar{\theta}+p\Bar{\phi}|^2+\\
    & |q\Bar{y}+r\Bar{z}+s\Bar{\alpha}+t\Bar{\beta}+u\Bar{\gamma}+v\Bar{\delta}+w\Bar{\theta}+x\Bar{\phi}|^2+|i\Bar{q}+j\Bar{r}+k\Bar{s}+l\Bar{t}+m\Bar{u}+n\Bar{v}+o\Bar{w}+p\Bar{x}|^2\\
    &<(1/8)
\end{aligned}$$ 
$E_{AC}:$
$$ \begin{aligned}
    & |a\Bar{e}+b\Bar{f}+c\Bar{g}+d\Bar{h}+i\Bar{m}+j\Bar{n}+k\Bar{o}+l\Bar{p}|^2+|a\Bar{q}+b\Bar{r}+c\bar{s}+d\Bar{t}+i\Bar{y}+j\Bar{z}+k\Bar{\alpha}+l\Bar{\beta}|^2+\\
    & |a\Bar{u}+b\Bar{v}+c\Bar{w}+d\Bar{x}+i\Bar{\gamma}+j\Bar{\delta}+k\Bar{\theta}+l\Bar{\phi}|^2+|e\Bar{u}+f\Bar{v}+g\Bar{w}+h\Bar{x}+m\Bar{\gamma}+n\Bar{\delta}+o\Bar{\theta}+p\Bar{\phi}|^2+\\
    & |q\Bar{u}+r\Bar{v}+s\Bar{w}+t\Bar{x}+y\Bar{\gamma}+z\Bar{\delta}+\alpha\Bar{\theta}+\beta\Bar{\phi}|^2+|e\Bar{q}+f\Bar{r}+g\Bar{s}+h\Bar{t}+m\Bar{y}+n\Bar{z}+o\Bar{\alpha}+p\Bar{\beta}|^2\\
    &<(1/8)
\end{aligned}$$
$E_{AD}:$
$$
\begin{aligned}
    & |a\Bar{c}+b\Bar{d}+e\Bar{g}+f\Bar{h}+i\Bar{k}+j\Bar{l}+m\Bar{o}+n\Bar{p}|^2+|a\Bar{q}+b\Bar{r}+e\Bar{u}+f\Bar{v}+i\Bar{y}+j\Bar{z}+m\Bar{\gamma}+n\Bar{\delta}|^2+\\
    & |a\Bar{s}+b\Bar{t}+e\Bar{w}+f\Bar{x}+i\Bar{\alpha}+j\Bar{\beta}+m\Bar{\theta}+n\Bar{\phi}|^2+|c\Bar{s}+d\Bar{t}+g\Bar{w}+h\Bar{x}+k\Bar{\alpha}+l\Bar{\beta}+o\Bar{\theta}+p\Bar{\phi}|^2+\\
    & |q\Bar{s}+r\Bar{t}+u\Bar{w}+v\Bar{x}+y\Bar{\alpha}+z\Bar{\beta}+\gamma\Bar{\theta}+\delta\Bar{\phi}|^2+|c\Bar{q}+d\Bar{r}+g\Bar{u}+h\Bar{v}+k\Bar{y}+l\Bar{z}+o\Bar{\gamma}+p\Bar{\delta}|^2\\
    &<(1/8)
\end{aligned}$$
$E_{AE}:$
$$
\begin{aligned}
    & |a\Bar{b}+c\Bar{d}+e\Bar{f}+g\Bar{h}+i\Bar{j}+k\Bar{l}+m\Bar{n}+o\Bar{p}|^2+|a\Bar{q}+c\Bar{s}+e\Bar{u}+g\Bar{w}+i\Bar{y}+k\Bar{\alpha}+m\Bar{\gamma}+o\Bar{\theta}|^2+ \\
    & |a\Bar{r}+c\Bar{t}+e\Bar{v}+g\Bar{x}+i\Bar{z}+k\Bar{\beta}+m\Bar{\delta}+o\Bar{\phi}|^2+|b\Bar{r}+d\Bar{t}+f\Bar{v}+h\Bar{x}+j\Bar{z}+l\Bar{\beta}+n\Bar{\delta}+p\Bar{\phi}|^2+\\
    & |q\Bar{r}+s\Bar{t}+u\Bar{v}+w\Bar{x}+y\Bar{z}+\alpha\Bar{\beta}+\gamma\Bar{\delta}+\theta\Bar{\phi}|^2+|b\Bar{q}+d\Bar{s}+f\Bar{u}+h\Bar{w}+j\Bar{y}+l\Bar{\alpha}+n\Bar{\gamma}+p\Bar{\theta}|^2\\
    &<(1/8)
\end{aligned} $$
$E_{BC}:$
$$
\begin{aligned}
    & |a\Bar{e}+b\Bar{f}+c\Bar{g}+d\Bar{h}+q\Bar{u}+r\Bar{v}+s\Bar{w}+t\Bar{x}|^2+|a\Bar{i}+b\Bar{j}+c\Bar{k}+d\Bar{l}+q\Bar{y}+r\Bar{z}+s\Bar{\alpha}+t\Bar{\beta}|^2+ \\
    & |a\Bar{m}+b\Bar{n}+c\Bar{o}+d\Bar{p}+q\Bar{\gamma}+r\Bar{\delta}+s\Bar{\theta}+t\Bar{\phi}|^2+|e\Bar{m}+f\Bar{n}+g\Bar{o}+h\Bar{p}+u\Bar{\gamma}+v\Bar{\delta}+w\Bar{\theta}+x\Bar{\phi}|^2+ \\
    & |i\Bar{m}+j\Bar{n}+k\Bar{o}+l\Bar{p}+y\Bar{\gamma}+z\Bar{\delta}+\alpha\Bar{\theta}+\beta\Bar{\phi}|^2+|e\Bar{i}+f\Bar{j}+g\Bar{k}+h\Bar{l}+u\Bar{y}+v\Bar{z}+w\Bar{\alpha}+x\Bar{\beta}|^2\\
    &<(1/8)
\end{aligned}$$
$E_{BD}:$
$$
\begin{aligned}
    & |a\Bar{c}+b\Bar{d}+e\Bar{g}+f\Bar{h}+q\Bar{s}+r\Bar{t}+u\Bar{w}+v\Bar{x}|^2+|a\Bar{i}+b\Bar{j}+e\Bar{m}+f\Bar{n}+q\Bar{y}+r\Bar{z}+u\Bar{\gamma}+v\Bar{\delta}|^2+\\
    & |a\Bar{k}+b\Bar{l}+e\Bar{o}+f\Bar{p}+q\Bar{\alpha}+r\Bar{\beta}+u\Bar{\theta}+v\Bar{\phi}|^2+|c\Bar{k}+d\Bar{l}+g\Bar{o}+h\Bar{p}+s\Bar{\alpha}+t\Bar{\beta}+w\Bar{\theta}+x\Bar{\phi}|^2+\\
    & |i\Bar{k}+j\Bar{l}+m\Bar{o}+n\Bar{p}+y\Bar{\alpha}+z\Bar{\beta}+\gamma\Bar{\theta}+\delta\Bar{\phi}|^2+|c\Bar{i}+d\Bar{j}+g\Bar{m}+h\Bar{n}+s\Bar{y}+t\Bar{z}+w\Bar{\gamma}+x\Bar{\delta}|^2\\
    &<(1/8)
\end{aligned}$$
$E_{BE}:$
$$
\begin{aligned}
    & |a\Bar{b}+c\Bar{d}+e\Bar{f}+g\Bar{h}+q\Bar{r}+s\Bar{t}+u\Bar{v}+w\Bar{x}|^2+|a\Bar{i}+c\Bar{k}+e\Bar{m}+g\Bar{o}+q\Bar{y}+s\Bar{\alpha}+u\Bar{\gamma}+w\Bar{\theta}|^2+\\
    & |a\Bar{j}+c\Bar{l}+e\Bar{n}+g\Bar{p}+q\Bar{z}+s\Bar{\beta}+u\Bar{\delta}+w\Bar{\phi}|^2+|b\Bar{j}+d\Bar{l}+f\Bar{n}+h\Bar{p}+r\Bar{z}+t\Bar{\beta}+v\Bar{\delta}+x\Bar{\phi}|^2+\\
    & |i\Bar{j}+k\Bar{l}+m\Bar{n}+o\Bar{p}+y\Bar{z}+\alpha\Bar{\beta}+\gamma\Bar{\delta}+\theta\Bar{\phi}|^2+|b\Bar{i}+d\Bar{k}+f\Bar{m}+h\Bar{o}+r\Bar{y}+t\Bar{\alpha}+v\Bar{\gamma}+x\Bar{\theta}|^2\\
    &<(1/8)
\end{aligned}$$
$E_{CD}:$
$$
\begin{aligned}
    & |a\Bar{c}+b\Bar{d}+i\Bar{k}+j\Bar{l}+q\Bar{s}+r\Bar{t}+y\Bar{\alpha}+z\Bar{\beta}|^2+|a\Bar{e}+b\Bar{f}+i\Bar{m}+j\Bar{n}+q\Bar{u}+r\Bar{v}+y\Bar{\gamma}+z\Bar{\delta}|^2+\\
    & |a\Bar{g}+b\Bar{h}+i\Bar{o}+j\Bar{p}+q\Bar{w}+r\Bar{x}+y\Bar{\theta}+z\Bar{\phi}|^2+|c\Bar{g}+d\Bar{h}+k\Bar{o}+l\Bar{p}+s\Bar{w}+t\Bar{x}+\alpha\Bar{\theta}+\beta\Bar{\phi}|^2+\\
    & |e\Bar{g}+f\Bar{h}+m\Bar{o}+n\Bar{p}+u\Bar{w}+v\Bar{x}+\gamma\Bar{\theta}+\delta\Bar{\phi}|^2+|c\Bar{e}+d\Bar{f}+k\Bar{m}+l\Bar{n}+s\Bar{u}+t\Bar{v}+\alpha\Bar{\gamma}+\beta\Bar{\delta}|^2\\
    &<(1/8)
\end{aligned}$$
$E_{CE}:$
$$
\begin{aligned}
    & |a\Bar{b}+c\Bar{d}+i\Bar{j}+k\Bar{l}+q\Bar{r}+s\Bar{t}+y\Bar{z}+\alpha\Bar{\beta}|^2+|a\Bar{e}+c\Bar{g}+i\Bar{m}+k\Bar{o}+q\Bar{u}+s\Bar{w}+y\Bar{\gamma}+\alpha\Bar{\theta}|^2+\\
    & |a\Bar{f}+c\Bar{h}+i\Bar{n}+k\Bar{p}+q\Bar{v}+s\Bar{x}+y\Bar{\delta}+\alpha\Bar{\phi}|^2+|b\Bar{f}+d\Bar{h}+j\Bar{n}+l\Bar{p}+r\Bar{v}+t\Bar{x}+z\Bar{\delta}+\beta\Bar{\phi}|^2+\\
    & |e\Bar{f}+g\Bar{h}+m\Bar{n}+o\Bar{p}+u\Bar{v}+w\Bar{x}+\gamma\Bar{\delta}+\theta\Bar{\phi}|^2+|b\Bar{e}+d\Bar{g}+j\Bar{m}+l\Bar{o}+r\Bar{u}+t\Bar{w}+z\Bar{\gamma}+\beta\Bar{\theta}|^2\\
    &<(1/8)
\end{aligned}$$
$E_{DE}:$
$$
\begin{aligned}
    & |a\Bar{b}+e\Bar{f}+i\Bar{j}+m\Bar{n}+q\Bar{r}+u\Bar{v}+y\Bar{z}+\gamma\Bar{\delta}|^2+|a\Bar{c}+e\Bar{g}+i\Bar{k}+m\Bar{o}+q\Bar{s}+u\Bar{w}+y\Bar{\alpha}+\gamma\Bar{\theta}|^2+\\
    & |a\Bar{d}+e\Bar{h}+i\Bar{l}+m\Bar{p}+q\Bar{t}+u\Bar{x}+y\Bar{\beta}+\gamma\Bar{\phi}|^2+|b\Bar{d}+f\Bar{h}+j\Bar{l}+n\Bar{p}+r\Bar{t}+v\Bar{x}+z\Bar{\beta}+\delta\Bar{\phi}|^2+\\
    & |c\Bar{d}+g\Bar{h}+k\Bar{l}+o\Bar{p}+s\Bar{t}+w\Bar{x}+\alpha\Bar{\beta}+\theta\Bar{\phi}|^2+|b\Bar{c}+f\Bar{g}+j\Bar{k}+n\Bar{o}+r\Bar{s}+v\Bar{w}+z\Bar{\alpha}+\delta\Bar{\theta}|^2\\
    &<(1/8)
\end{aligned}$$

\end{document}